\documentstyle[prb,eqsecnum,aps,twocolumn]{revtex}

\begin{document}
\draft
\twocolumn[\hsize\textwidth\columnwidth\hsize\csname @twocolumnfalse\endcsname

\title{
Impurity scattering and transport of fractional Quantum Hall edge states }

\author{C.L. Kane}

\address{Department of Physics, University of Pennsylvania\\
Philadelphia, Pennsylvania 19104
}
\author{Matthew P.A. Fisher}
\address{Institute for Theoretical Physics, University of California\\
Santa Barbara, CA 93106
}

\date{\today}
\maketitle

\begin{abstract}
We study the effects of impurity scattering on the low energy edge state
dynamics for a broad
class of quantum Hall fluids at filling factor $\nu =n/(np+1)$,
for integer $n$ and even integer $p$.
When $p$ is positive all $n$ of the edge modes are expected
to move in the same direction, whereas for negative $p$
one mode moves in a direction opposite to the other $n-1$ modes.
Using a
chiral-Luttinger model to describe the edge
channels, we show that for an ideal edge when $p$ is negative,
a non-quantized and non-universal Hall conductance is predicted.
The non-quantized conductance is
associated with an absence of equilibration between
the $n$ edge channels.  To explain the robust
experimental Hall quantization, it is thus necessary
to incorporate impurity scattering into the model, to allow for edge
equilibration.  A perturbative analysis reveals that edge
impurity scattering is relevant and will modify the
low energy edge dynamics.  We describe a non-perturbative solution
for the random $n-$channel edge, which reveals the existence of a new
disorder-dominated phase, characterized by a stable
zero temperature renormalization group fixed point.
The phase consists of a single propagating charge mode,
which gives a quantized Hall conductance, and $n-1$ neutral modes.
The neutral modes all propagate at the same speed, and manifest
an exact SU(n) symmetry.  At finite temperatures the
SU(n) symmetry is broken
and the neutral modes decay with a finite rate which
varies as $T^2$
at low temperatures.
Various experimental predictions and implications which
follow from the exact solution are described in detail,
focusing on tunneling experiments through point contacts.

\end{abstract}

\pacs{PACS numbers:  72.10.-d   73.20.Dx}
]

\section{Introduction}

It was over a quarter of a century ago that pioneering theoretical work on
one-dimensional
interacting electron gases demonstrated the profound effects that
electron-electron interactions can have in low-dimensional quantum systems
\cite{lutt1,lutt2,lutt3}.
Specifically, it was found that even weak repulsive interactions
de-stablize a Fermi-liquid description of a one-dimensional
electron gas.  Some years later\cite{Haldane1}, Duncan Haldane
coined the term "Luttinger liquid", to describe
the generic state of a one-dimensional
interacting electron gas.  Since then
it has become possible to fabricate one dimensional
electron gases in semiconductors\cite{Wind,wire}, by clever lithography on
semiconductor heterostructures.  Unfortunately, searches for
non-Fermi liquid properties in such one-dimensional quantum wires has been
impeded by spurious impurities\cite{Webbp}, which tend to localize the
electrons.
It has recently been emphasized\cite{review,Wen1,MacDonald,Moon}, though, that
the Quantum Hall
effect might serve as an alternate
arena to study one-dimensional electron gases.
In the presence of a strong external magnetic field
a two-dimensional electron gas forms an incompressible quantum Hall
fluid\cite{Prange-Girvin}
and the current which flows
is confined to the edges.  These current carrying edge states provide a unique
laboratory for the study of "ideal" one-dimensional systems.

Indeed, a key feature of Quantum Hall edge states is their resilience in the
presence of disorder\cite{Halperin1}.
For example, electrons on the edge
of a quantum Hall state at filling factor $\nu=1$,
which corresponds to a full Landau level,  are completely
insensitive to the presence of disorder on the edge.  This is
because edge state electrons propagate
in only one direction, and hence cannot be backscattered by
random impurities.  There is no localization, and
the only effect of disorder is to give the electrons
an unimportant forward scattering phase shift.
In fact, the very quantization of the Hall conductance in the integer
effect can be understood simply in the framework of the Landauer-
Buttiker theory\cite{Buttiker}, which relates the quantized conductance
to the perfect transmission of free electron edge states.

When there are multiple edge channels, such as for
the integer quantum Hall effect at $n$ full Landau levels, $\nu=n$,
disorder plays a more important
role by causing the scattering of electrons
between the different channels\cite{Wen2}.
However, in this case the channels all propagate in the same
direction, so that, as indicated in figure 1a,
the net current is not altered by the scattering
events.  The total transmission and resulting conductance is still
quantized.  Nonetheless,
it has been possible to probe such inter-channel scattering by
selectively "feeding" different edge modes from
contacts at different
chemical potentials,
and then examining the resulting equilibration
\cite{Alphenaar,Kouwenhoven}.

In the fractional quantum Hall effect, the free electron edge state
theories can no longer be applied.
Recently, however, an alternative description has been
proposed by Wen \cite{Wen1} in which the
edge states in the fractional
quantum Hall effect are described by a chiral Luttinger liquid
model.  In particular, for the Laughlin
states, such as $\nu=1/3$, the edge states consist of a single
branch, and are described by a single channel chiral Luttinger liquid.
As in the case of $\nu=1$, edge disorder is expected to play
no role in these states.  Recent experiments on tunneling
between edge states at a point contact in the
$\nu=1/3$ quantum Hall effect support the chiral Luttinger liquid
model\cite{Webb}.

For hierarchical quantum Hall states there will
be many branches of edge excitations\cite{Wen1}.  In general, at the
n'th level of the Haldane-Halperin hierarchy\cite{HaldaneHalperin},
the appropriate
description is an n channel chiral Luttinger liquid.
In Wen's theory the universal properties of the
bulk Hall fluid determine the direction of propagations
of the edge modes.  In contrast to
the integer quantum Hall effect, there is a class of
fractional Hall states for which the n edge modes
are not all moving in the same direction\cite{Wen1,MacDonald}.
Unfortunately, as we show in detail below, in these cases the theory predicts
a value of the conductance which is not correctly quantized,
depending on non-universal interaction parameters at the edge.

In a recent paper with Joe Polchinski\cite{us23}
we argued that for this class of fractional Hall
states it is absolutely crucial to include impurity scattering at the edge.
Such scattering allows for charge transfer between
channels moving in opposite directions, as shown in figure
1(a), and can modify the conductance.
Specifically, we studied the effects of such inter channel
impurity scattering
at the edge of a $\nu=2/3$ quantum Hall state\cite{us23}.
In the absence of impurities, the simplest model\cite{Wen1,MacDonald} of a
$\nu=2/3$ edge consists of two charged
modes:
one with conductance $e^2/h$ and another with conductance $(1/3)e^2/h$
which moves in the opposite direction.  We found that
even weak interchannel impurity scattering is
relevant, and at low energies the edge is described by a new
disorder
dominated phase.  An exact solution in this phase
revealed the presence of a single charged mode, which gave
the correct quantized conductance of $2/3(e^2/h)$, and
a neutral mode which propagates in the opposite direction.  The neutral
mode was shown to possess an exact
SU(2) symmetry.

In this article we elaborate significantly on
the above results and generalize them
to fractional quantum Hall states at higher levels of the hierarchy.
Specifically, we consider quantum Hall states
at filling factors $\nu = n/(np+1)$, with $p$ an even integer
and $n$ an arbitrary positive integer.  Within
Jain's hierarchical construction \cite{Jain},
these states can be achieved by attaching flux tubes with
$p$ flux quanta to each electron, and putting the resulting
composite fermions in $n$ full Landau levels.
With this convention, $p$ can be a negative even integer,
in which case the filling factor is $-\nu$.
For this broad class
of quantum Hall fluids, we
find the presence of edge impurity scattering drives the edge modes to a
new fixed point in which the charge and neutral sectors decouple
at low energies.  More specifically, the edge fixed point is
characterized
by a single propagating charged mode
with conductance $|\nu| e^2/h$
and $n-1$ neutral modes.  The $n-1$ neutral modes will be
shown to have an exact SU(n) symmetry, implying
that they all move at the same velocity.
The direction of propagations of the neutral modes with respect to
the charge mode is determined by the sign of $p$, moving in a direction
opposite to the charge mode for $p$ negative.

Since our initial Hamiltonian for the edge modes has
only one conserved U(1) charge,
the physical electric charge, the presence of the additional
$n-1$ propagating neutral modes is quite surprising.
However, the fixed point to which the system scales
at low energies has much higher symmetry
- an exact U(1)xSU(n) symmetry - than
the original Hamiltonian.
Indeed, it is the presence of the
SU(n) symmetry at the attractive fixed point which leads to the existence of
the $n-1$ additional neutral modes.

Since the fixed point is a zero temperature fixed point,
the  SU(n) symmetry is broken at finite temperatures.
It follows that at $T \ne 0$
the neutral modes are not conserved and will decay with a lifetime
$\tau_\sigma$, or equivalently a finite decay length
$\ell_\sigma = v_\sigma \tau_\sigma$,
where $v_\sigma$ is the velocity of the neutral modes.
By analyzing the leading irrelevant operators
which control the flows into the zero temperature fixed point, we will show
that
the decay rate vanishes algebraically at zero
temperature:
\begin{equation}
{1 \over {\tau_\sigma}} \propto T^2.
\end{equation}
In contrast,
the charge mode cannot decay, even at finite temperature, since
electric charge is always conserved.
However, due to irrelevant operators which couple
the charge and neutral sectors, the charge mode can scatter
off the neutral modes.  This leads to a charge mode
which propagates with a dispersion
$\omega = v_\rho q + i D q^2$, with a "diffusion" constant $D$ which is
temperature
independent at low temperatures.  This implies a diffusive
spreading of a charge pulse as it propagates along an edge.

On length scales longer than $\ell_\sigma$, it is appropriate to
adopt a ``hydrodynamic" description of the edge propagation,
in which
there is only a single propagating mode associated
with the conserved electric charge.
However, this hydrodynamic
picture leaves out important low temperature physics,
which can be accessed via inter-edge tunneling.  We shall return
to this point in section IV.

The existence of an attractive zero temperature fixed point with
higher symmetry than the underlying Hamiltonian is reminiscent of
Fermi Liquid theory.  The zero
temperature Fermi liquid fixed point has,
in addition to conserved electric charge,  an infinity of conserved charges
(and hence an infinity of U(1) symmetries)
associated with each point on the Fermi surface.
This is the symmetry responsible for the
quasiparticle excitations.
At
finite temperatures this symmetry is broken,
leading to a finite scattering lifetime for the quasiparticles,
proportional to $T^{-2}$.
Since the total electric charge is conserved, there remains a propagating
zero sound mode in a Fermi-liquid, which does not decay.
It is amusing that we find a scattering rate for the neutral edge
excitations, Eqn.(1.1),  which vanishes with the same power
of temperature - $T^2$ - as the quasiparticles in a Fermi liquid.

At low temperatures
the restoration of the full SU(n) symmetry at the edge of a random
$\nu=n/(np+1)$ Hall state has important experimental
consequences.  When $p$ is negative, and the
neutral modes travel in the opposite direction to the charge mode,
the very quantization of the Hall conductance rests on this
symmetry.
As we show explicitly below, in the absence of edge randomness which
equilibrates
the edge modes,
a non-universal value of the Hall conductance is
predicted\cite{us23} for $p<0$.

The presence of the SU(n) edge symmetry also
implies universal values for the scaling dimensions of the
edge tunneling operators.  These scaling dimensions
are experimentally accessible, by measuring the temperature
dependence of the tunneling conductance through a
point contact\cite{Webb,Moon,us1}.  Our central prediction is that
when $p$ is negative, the conductance through a point contact
in a $\nu = n/(np+1)$ Hall fluid should vanish as
\begin{equation}
G(T) \propto  T^\alpha,
\end{equation}
where
\begin{equation}
\alpha = 2|p|-(4/n) .
\end{equation}
For the $p=-2$ sequence, the predicted exponents are
displayed in table 1.  The exponents approach
$\alpha = 4$ as $\nu$ approaches 1/2.  For non-negative $p$
an exponent $\alpha = 2p$ is predicted.

A measurement of temperature exponents consistent with these
would
give indirect evidence of the neutral modes, since the
electron which tunnels through the point contact is "built"
from a superposition of the charge and neutral edge modes.
The neutral modes should be measurable more directly, though,
via time domain experiments, which we discuss below.
In this way, one might be able to measure directly
the temperature dependent decay rate of the neutral mode -
roughly analogous to a direct measurement of a decaying
Fermi-liquid quasiparticle.

The outline of our paper is as follows.  In Section II we
describe the model for an impurity free quantum Hall edge
at filling $\nu=n(np+1)$, which consists of an n-channel Chiral
Luttinger liquid.  We split the model into two
pieces, denoted $S_0$ and $S_1$,  and show that the first piece can be
conveniently decoupled
into a charge sector and a neutral sector with $n-1$ modes.
We then demonstrate that
$S_0$ possesses an exact U(1)xSU(n)
symmetry.  This symmetry is not respected by $S_1$, however.
In Section III we consider the addition of the most general
random impurity scattering terms.  Although these random terms break the
U(1)xSU(n)
symmetry of $S_0$, we show in Section IIIa that provided $S_1$ is ignored the
random
model can be solved exactly.  In terms of new fields, the exact solution
reveals an exact U(1)xSU(n) symmetry.  In Section IIIb we show that
the exact solution is perturbatively stable in the presence
of non-zero $S_1$.  The effects of small non-zero temperatures
are considered in Section IIIc.  In Section IV we use the exact
solution of the random edge to calculate the scaling dimension
of edge tunneling operators, which are relevant to experiments
on tunneling through a point contact.
Section V is devoted
to specific experimental predictions, and a more general discussion
of our central results.

\section{THE CLEAN EDGE}

\subsection{The Model}

The topological order of a quantum Hall state in the n'th level of the
hierarchy is characterized by a symmetric $n\times n$ matrix $K$.
The low energy physics of a
hierarchical quantum Hall state
may be described by $n$ gauge fields with an effective action
\cite{Read,WenZee},
\begin{equation}
S_{\rm bulk} = {i \over {4 \pi}} \int a_\mu ^i K_{ij}
\epsilon_{\mu \nu \lambda} \partial_\nu a_\lambda ^j.
\end{equation}
We use the "symmetric" basis in which
the electron 3-current is given by
\begin{equation}
j_\mu =  \sum_i \epsilon_{\mu\nu\lambda} \partial_\nu a^i_\lambda /2\pi  .
\end{equation}
In this basis the filling factor is given by
\begin{equation}
\nu = \sum_{ij} K^{-1}_{ij}   .
\end{equation}
For the quantum Hall states at filling
$\nu =n/(np+1)$, in both the Haldane/Halperin hierarchy and
in the Jain construction, the $K$ matrix is given explicitly by
\begin{equation}
K_{ij} = \delta_{ij} + p  .
\end{equation}
The $K$ matrix characterizes the charge and statistics of the bulk
quasiparticle excitations.  Specifically,
the quasiparticles are labeled by a set of integers $m_j$, with
$j=1,2,..,n$, and in the symmetric basis have a charge (in units of the
electron charge)
\begin{equation}
Q =  \sum_{ij} m_i K^{-1}_{ij}
\end{equation}
and statistics angle
\begin{equation}
{\Theta\over\pi} =  \sum_{ij} m_i K^{-1}_{ij} m_j   .
\end{equation}
In this approach, all of the universal properties of the bulk quantum Hall
state
follow directly from the $K$ matrix.

It is worth emphasizing the implicit assumptions which were needed to
arrive at the simple form (2.1).  These can be perhaps most
easily understood in terms of the Ginzburg-Landau description\cite{GL} of
the Hall effect.  For $\nu=n/m$ with $m$ odd, $n$ electrons bind
with $m$ vortices forming a "molecule" with bosonic statistics.
At the magic rational filling factor $\nu=n/m$, all of the vortices induced by
the
magnetic field are accommodated in this way.  The electron/vortex
composites can then Bose condense, leading to the quantum Hall effect.
The effective action (2.1) describes the long wavelength density fluctuations
of this condensed fluid.  The bulk quasiparticle excitations,
referred to above, are essentially excitations involving breaking
apart the electron/vortex composites.  Although the $K$ matrix
determines the charge and statistics of these quasiparticles,
the energy gap for their creation is not specified by
the effective action (2.1).  Provided the temperature is well
below these energy gaps, the effective action (2.1) provides
an adequate description.  However, at filling factors away from $\nu=n/m$,
there will be some residual vortices, and the electron/vortex composites
can only condense if these residual vortices are pinned
and localized by bulk impurities.  In this case, there will
be many low energy, but spatially localized, excitations involving
re-arranging the positions of these vortices.  The effective action (2.1)
can presumably still be used to extract transport properties, though,
since at low temperatures the localized vortices will not contribute
significantly to the transport.  (This is not the case for other physical
properties such as the electronic specific heat.)
Although the quasiparticle excitations are not important
at low $T$ in the bulk, they play a crucial role at the edge.
At the edge their gap vanishes and they
form the edge states, which we next discuss.

As shown by Wen\cite{Wen1},
the edge excitations may be described by eliminating the
bulk degrees of freedom from (2.1).
Upon integration over $a^i_\tau$, a constraint on the density
fluctuations in the bulk is imposed:  $\vec\nabla \times \vec a^i=0$,
for all $i=1,2,...,n$.  Here a vector refers to the two spatial
components.  Scalar fields can then be introduced to solve these
constraints, $\vec a^i = \vec\nabla \phi_i$, one for each gauge
field.  The edge excitations are then described in terms of these
scalar fields. The appropriate effective action at the
edge can then be written as
$S = S_0 + S_1$ with
\begin{equation}
S_0  =
 \int dx d\tau {1\over {4\pi}} [ \sum_{ij} (\partial_x\phi_i)K_{ij}
(i\partial_\tau \phi_j ) + v \sum_i (\partial_x \phi_i )^2 ]
\end{equation}
and
\begin{equation}
S_1 = \int dx d\tau {1\over {4\pi}}
\sum_{ij} V_{ij} \partial_x \phi_i \partial_x \phi_j
\end{equation}
with $\sum_i V_{ii} = 0$.
Here $x$ is a one-dimensional spatial coordinate which runs along the edge,
and $\tau$ is imaginary time.
In addition to the $K$ term, whose form is determined solely
from the bulk physics, we also have interaction terms of the form
$\partial_x \phi_i \partial_x \phi_j$.
These interaction strengths are non-universal, and depend on the form of the
edge confining potential
and the details of the electron-electron interactions (which we assume
here to be short ranged, screened by a ground plane).
For later convenience we have split these interaction terms into a constant
velocity piece,
$v$ in $S_0$, and a traceless velocity matrix $V_{ij}$ in $S_1$.

It follows from Eqn. (2.2) that the one-dimensional
electron charge density along the edge is given by
\begin{equation}
\rho(x) = {1 \over {2 \pi}} \sum_{i=1}^n \partial_x  \phi_i  .
\end{equation}
Operators which create charge at the edge can be deduced by noting that the
momentum conjugate to the
fields is $\Pi_i = (1/2\pi) K_{ij} \partial_x \phi_j$.
Thus an operator of the form $\exp i\phi_i(x)$, which can be expressed
as a spatial integral over the conjugate momenta,
creates "instantons" in the boson fields $\phi_j$ at position $x$.
These instantons carry electron charge, as can be seen from (2.9).
Specifically, the general edge creation operator
\begin{equation}
\hat{T}(x) = e^{i\sum_{j=1}^n m_j \phi_j(x)}
\end{equation}
for arbitrary integers $m_j$, creates an edge excitation at $x$ with charge
$Q$ given in (2.5).

\subsection{Absence of edge equilibration}

The beautiful feature of the effective action (2.7)-(2.8)
is it's simplicity:  It is quadratic in the boson fields,
and all physical quantities can thus be easily computed.
Unfortunately, when $p$ is negative, the results are
in serious conflict with experiment.
The most worrisome conflict involves the Hall conductance itself\cite{us23},
which we find is not given by the quantized value $|\nu| e^2/h$\cite{coulomb}.

The difficulty occurs when all of the edge modes do not propagate
in the same direction.  As shown in Appendix A, the sign of
the eigenvalues of the $K$ matrix determine the direction of propagation of
the eigenmodes.  We show in Section IIc below that for $\nu = n/(np + 1)$ the
$K$ matrix
has $n-1$ degenerate eigenvalues equal to one,
and one eigenvalue equal to (1+np).
Thus, when $p$ is negative, there is one mode which moves in
a direction opposite to the other $n-1$ modes.

To show that the conductance is non-universal for negative $p$
and to gain a physical understanding for why this is, it is useful to
generalize Landauer-Buttiker
transport theory to that of an interacting
Luttinger liquid.
To this end, consider an edge state
which flows between two reservoirs
which are in equilibrium
at different chemical potentials (see Figure 2).  We model the reservoirs
by considering an infinite edge, in which the "sample"
resides between $x_L$ and $x_R$.  The left and right
reservoirs are then defined for $x<x_L$ and $x>x_R$ respectively.
We suppose that the system is driven from equilibrium by an
electrostatic potential $eV(x)$, which
couples to the edge charge density $\rho(x)$, and is a constant
$eV_{L(R)}$ in the left (right) reservoir.
The underlying physical assumption of this approach is that
the edge states which emanate from a given reservoir are in
equilibrium at the chemical potential of that reservoir.

Since the edge current operator is linear in the boson
fields,
\begin{equation}
\hat I_{\rm edge} = e {1 \over {2\pi}}  \sum_{j=1}^n \dot\phi_j,
\end{equation}
the edge current at a point $x$
which flows in linear response to $V(x')$ may be computed directly.
Specifically,
\begin{equation}
I_{\rm edge}(x) =  \int dx' D^R(x-x',\omega\rightarrow 0)  V(x'),
\end{equation}
where the retarded response function is given by
\begin{eqnarray}
\nonumber
D^R && (x-x',\omega) =  -i \int_{-\infty}^0 dt e^{-i\omega t} \\
 && \sum_{i,j} {e^2\over {(2\pi)^2 \hbar}}
< [ \dot\phi_i(x,0),\partial_{x'} \phi_j(x',t)] >.
\end{eqnarray}

Consider first the simple case of a single channel
edge, such as $\nu = 1/m$, described by the action (2.7) with
$K_{11} = \eta m$.  Here $\eta = \pm 1$ determines the direction of
edge propagation.  Using (2.7) the
response function
(2.13) may be readily computed by analytically continuing the
imaginary time response function
\begin{equation}
D(x-x',\omega_n) = - {1\over m} {e^2\over h}
\sum_q e^{iq(x-x')} {q \omega_n \over {q(\eta i\omega_n - v q)}}
\end{equation}
to real frequencies, $i\omega_n \rightarrow \omega+ i\epsilon$.
We then find
\begin{equation}
D^R(x-x',\omega) = {1\over m} {e^2\over h} \theta\left(\eta(x-x')\right)
\ {i\eta \omega\over v} e^{i \eta (\omega+i\epsilon)(x-x')/v}.
\end{equation}
Note the presence of the $\theta$ function which shows that the
current at $x$ depends only on the voltages at positions $x'$
that are ``upstream" of $x$.  This reflects the chiral nature of
the edge state propagation.  In the limit $\omega\rightarrow 0$,
the integral in (2.12) will be dominated by values of $x'$ that
are deep into the ``upstream" reservoir.  Thus, for
$\eta=+1$, which corresponds to
an edge which propagates from left to right,
the current is
\begin{equation}
I_{\rm edge}  = {1\over m} {e^2\over h} V_L .
\end{equation}
The two terminal conductance of a Hall bar in the $\nu=1/m$ state
follows if we consider, in addition, the opposite edge which
emanates from the right reservoir and contributes a current
$-1/m (e^2/h) V_R$.  The net current is thus $I = G (V_L - V_R)$,
with an appropriately quantized two terminal conductance: $G= (1/m) (e^2/h)$.

This approach can easily be generalized to the hierarchical quantum
Hall states, which have multiple channels.  However, the situation
is more complicated if channels on a given edge move in both
directions.
In figure 2 we consider a two channel example in which the top edge
contains two modes which propagate in opposite directions.  Clearly,
the current on the top edge will depend on the voltages in both
reservoirs.
In appendix A we show
that in general the edge current may be written
\begin{equation}
I_{\rm edge} =  {e^2\over h}\left( g_+ V_L - g_- V_R\right),
\end{equation}
where $g_+$  ($g_-$) is the total dimensionless conductance from all
right (left) moving channels.
When all of the channels move in the same direction either
$g_+$ or $g_-$ will be equal to zero.  However, for $p<0$, when channels
move in both directions, they will both be positive.

The two terminal conductance then follows by considering the
other edge, which carries a current $g_- V_L - g_+ V_R$.
Thus we find
\begin{equation}
G = {e^2\over h} ( g_+ + g_- ) .
\end{equation}
Notice that the conductances of each mode add in parallel, irrespective
of their direction of propagation.
In appendix A we explicitly compute $g_+$ and
$g_-$ using the effective action (2.7)-(2.8), and show that they
are non universal, depending on
the interaction strengths $V_{ij}$ in (2.8).  However, the combination
$g_+ - g_- = \nu$ is shown to be universal.
Thus we see that if all channels move in the same direction
the two terminal conductance has the quantized value
$G= |\nu| e^2/h$.  However, when there are channels moving
in both directions, the two terminal conductance will be
non universal, and will in
general be larger than $|\nu| e^2/h$.

It is straightforward
to generalize the above approach, based
on the right/left conductances, $g_\pm$, to compute the conductance
measured in a
four terminal geometry.  In particular, we find that the
four terminal
Hall conductance is given by
\begin{equation}
G_H = {e^2\over h}{g_+^2 + g_-^2\over {g_+ - g_-}}.
\end{equation}
Thus it is only when all channels propagate
in the same direction that $G_H$ is universal and equal to $\nu e^2/ h$.

In appendix A we also show
that the scaling dimensions of tunneling operators
are similarly universal only when all channels move in the same
direction.  These scaling dimensions
enter into experimentally accessible
quantities, such as the temperature
dependence of tunneling through a point contact.
We will discuss this point in more detail in section IV.

We thus see that when $p$ is negative, the Luttinger
edge model, (2.7)-(2.8), predicts
a two terminal and Hall conductance which is not quantized,
in glaring contradiction with experiment.  Clearly
some important physics must be absent from the simple effective action
(2.7).
A clue can be seen from Figure 2, where it is clear that
in a transport
situation, right moving edge modes are in
equilibrium with the left reservoir, and left movers
in equilibrium with the right reservoir.
Thus in the presence of a non-zero source-to-drain voltage,
opposite moving edge modes on a given edge
will be out of equilibrium
with one another.

But since these modes are in close
proximity, what stops them from
equilibrating?
In the effective action (2.7)-(2.8) there are simply no terms
which transfer charge between the different edge modes,
to allow for possible equilibration.
But surely in real experimental systems there
will be equilibration processes present.
A constraint is that charge transfer between edge modes must
conserve momentum along the edge.  However, different
edge modes will have different momenta - the gauge invariant
momentum difference between two modes being proportional to
the magnetic flux threading the space between them.
Since in equilibrium the different edge modes are at the same energy,
processes which transfer charge between two edge modes,
with the emission of phonons or photons to take up
the momentum, will not conserve overall energy.
These processes are thus forbidden.

However, if there are impurities near the edge, as there will
be in any real sample, the momentum of the edge modes
need not be conserved.  Momentum can be transferred to the
center of mass of the crystal sample, through the impurities.
Thus a disordered edge with impurity scattering will allow for possible
equilibration between the different edge modes.
In Section III we study the effect of impurity edge scattering.
Before doing so, it is useful to first establish the
existence of a special SU(n) symmetry in the action $S_0$.
This symmetry will be crucial in arriving at a solution of the
disordered edge.

\subsection{SU(n) Symmetry of $S_0$}

It has been known for some time that the structure of the
$K$ matrix at filling $\nu = n/(np+1)$ implies that equation (2.1) possesses
a hidden SU(n) symmetry\cite{Read,Zeesun}.
This is most readily seen for the special case $p=0$, which corresponds
to the integer quantum Hall effect with filling
$\nu =n$.  However, additional non-universal
terms should be added to (2.1) (for example terms with
two or more derivatives), and these terms will not
respect the SU(n) symmetry.  So in general the
SU(n) symmetry is not expected to be manifest in the bulk.
Again, this can be seen clearly in the integer
quantum Hall effect ($p=0$), where the quasi-hole excitation energies
in the n-full levels will not be the same.

The SU(n) symmetry implied by the form of the $K$ matrix is also
manifest at the edge.  For the integer Hall effect ($p=0$)
the SU(n) symmetry is apparent in the edge action $S_0$
which corresponds to n identical channels of
chiral fermions.
However, as in the bulk, this symmetry will in general be
broken by non-universal terms, for example the velocity matrix $V_{ij}$
in $S_1$, which has no special symmetry properties.

A random edge potential will introduce additional terms which also break
the SU(n) symmetry.  However, the very presence of
these random terms drives the edge at low energies into a phase
in which the SU(n) symmetry is restored.
This will also be the case for non-zero $p$.

We now
show that the action $S_0$
has an SU(n) symmetry even for non-zero p.
This will be accomplished via a transformation
which decouples the charge degree of freedom, described by
$\phi_\rho = \sum_i \phi_i$, from the remaining neutral degrees
of freedom.
The neutral sector can then be mapped onto the neutral sector of a
$\nu=n$ edge, which is described by SU(n) chiral fermions.
It is useful to first introduce some
SU(n) notation.
We denote the $n-1$ diagonal generators as $D^m$ with $m=1,2,...n-1$.
To be specific we take $D^m$ to be $n\times n$
diagonal matrices with m ones along the diagonal,
starting from the upper left,
with the next diagonal element being -m, to make the matrix traceless.
The matrices are then divided by a normalization factor
$\sqrt{m^2 + m}$ to make
${\rm tr}(D^mD^m)=1$.  We denote the $n(n-1)$ non-diagonal generators of SU(n)
as $R^{ij}$, ($i\neq j =1,2,...n)$ , which have a single non-zero
matrix element, the (ij)-element, equal to one.

The decoupling of the charge and neutral sectors
may be performed by defining new fields
\begin{equation}
\Phi_i = O_{ij} \phi_j
\end{equation}
where the matrix $O_{ij}$ is an orthogonal transformation,
$O^T O=1$, given by,
\begin{equation}
O_{ij} = D^i_{jj}
\end{equation}
for $ i=1,2,..,n-1$ (no sum on $j$) and
\begin{equation}
O_{nj} = 1/\sqrt n   .
\end{equation}
It can be readily checked that this transformation diagonalizes
the matrix $K$, giving for
$\tilde{K} = OKO^{-1}$ a diagonal matrix of the form:
$\tilde K_{ij} = \delta_{ij} (1+ np \delta_{in})$.
Upon defining a charge field $\phi_\rho = \sqrt{n}\Phi_n = \sum_i \phi_i$,
so  that the total edge density is given by
$\rho = \partial_x\phi_\rho/2\pi$,
the action $S_0$ is seen to de-couple into a charge
and neutral sector, $S_0=S_\rho + S_\sigma$, with
\begin{equation}
S_\rho = \int dx d\tau {1 \over {4\pi}} [ {1 \over \nu} i\partial_\tau
\phi_\rho
\partial_x \phi_\rho + v (\partial_x \phi_\rho )^2 ]
\end{equation}
and
\begin{equation}
S_\sigma = \int dx d\tau {1 \over {4\pi}} \sum_{i=1}^{n-1}
\partial_x \Phi_i ( i \partial_\tau + v \partial_x ) \Phi_i  .
\end{equation}
Notice that when the even integer $p$ is negative,
$\nu=n/(np+1)$ is negative, and
the charge mode moves in a direction opposite to the ($n-1$)-neutral modes.

In order to make the SU(n) symmetry more explicit, we map $S_\sigma$
onto the neutral sector of SU(n) fermions.  To accomplish this we
introduce an additional auxiliary field, $\tilde\Phi_n$,
which has an action identical to each of the neutral modes
in $S_\sigma$.  Upon adding this action to it one has
\begin{equation}
S_\sigma \rightarrow  \int dx d\tau {1 \over {4\pi}} \sum_{i=1}^{n}
\partial_x \tilde\Phi_i ( i \partial_\tau + v \partial_x ) \tilde\Phi_i
\end{equation}
where we have defined $\tilde\Phi_i = \Phi_i$ for $i=1,,2,...,n-1$.
It is finally convenient to rotate back, via
$\tilde\phi_i = O_{ji} \tilde\Phi_j$ which leaves
the form for $S_\sigma$ unchanged.
The final step is to fermionize the resulting boson fields
\begin{equation}
\psi_i \propto e^{i\tilde\phi_i} .
\end{equation}
In this way the free action can finally be expressed as
\begin{equation}
S_0 = S_\rho + \int dx d\tau \psi^\dagger (\partial_\tau
-iv \partial_x ) \psi
\end{equation}
where $\psi$ here denotes an n-component Fermion field.
The SU(n) symmetry of the neutral sector is thus manifest.
The U(1) charge sector of the above chiral fermions is precisely
the auxiliary field
$\tilde\Phi_n$, introduced above.  This field does not enter into any physical
quantities, but allows for the above convenient (fermion) representation
of the SU(n) symmetry in the neutral sector.

\section{THE RANDOM EDGE}

Having established the inadequacies of the
clean edge, described by the effective action (2.7)-(2.8), we
consider now the effects of edge impurity scattering,
which allows for inter-channel equilibration.
With disorder present we will show
that the low temperature physics is described by
a new random fixed point, which can be solved exactly.

In the presence of impurity scattering there are many
different types of random edge operators which
can be added to the pure action $S_0 + S_1$ given in
(2.7) and (2.8).
Here we focus on those which are most relevant.
The simplest random terms will take the form:
\begin{equation}
{1\over{2\pi}}
\int dx d\tau \sum_i \mu_i (x) \partial_x \phi_i
\end{equation}
where the $\mu_i$ are spatially dependent random potentials,
which couple to the density in each mode.
These terms are unimportant, however,  since they can be eliminated
from the action via a transformation,
\begin{equation}
\phi_i(x) \rightarrow \phi_i(x)  + \int_{-\infty}^x dx' \sum_j M_{ij}\mu_j
(x'),
\end{equation}
where $M_{ij}^{-1} = v \delta_{ij} + V_{ij}$.

More important are random terms which tunnel quasiparticles
between the n edge modes, allowing for equilibration.
The most relevant operator which tunnels
charge between channel i and j is given by
$\exp i(\phi_i - \phi_j)$.  A random impurity potential will give rise to
terms in the action of the form
\begin{equation}
S_{\rm random} = \int dx d\tau \sum_{i>j} \left[ \xi_{ij}(x)
e^{i(\phi_i - \phi_j)} + h.c. \right]
\end{equation}
where $\xi_{ij}(x)$ are spatially random tunneling amplitudes
between edge modes $i$ and $j$.  These amplitudes
are complex because the different edge channels have different momenta.
Indeed, for a clean edge
the tunneling amplitude would oscillate, as $\exp ik_{ij}x$, where
the gauge invariant momentum difference,
$k_{ij}$, is proportional to the magnetic flux per unit length enclosed
between the two channels.   This would
be ineffective
at equilibrating -  however, with impurity
scattering present, momentum of the edge modes is not conserved,
and equilibration can take place.

Since the operators entering into
$S_{\rm random}$ are non-linear in the boson fields,
the full random model appears rather intractable.
One approach is to study the effects of
the random potential $\xi_{ij}(x)$ in perturbation theory about the
free theory, $S_0+S_1$.
This is problematic, however, because the perturbation theory is
divergent at low energies.
One can nevertheless define a perturbative renormalization
group transformation in powers of the variance, $W_{ij}$,
defined via $[\xi_{ij}^*(x) \xi_{ij}(0) ]_{ens} = W_{ij} \delta(x)$,
where the square brackets denote an ensemble average over realizations
of the disorder\cite{Giamarchi}.  The leading order renormalization group flow
equations
take the form
\begin{equation}
{\partial W_{ij}\over{\partial\ell}} = (3-2\Delta_{ij})W_{ij}
\end{equation}
where $\Delta_{ij}$ is the scaling dimension of the operator
$\hat O_{ij} = \exp(i(\phi_i - \phi_j ))$ evaluated in the free theory,
defined as $<O^\dagger (\tau) O(\tau=0)> \sim \tau^{-2\Delta}$.
These scaling dimensions are computed explicitly in appendix A.
At the SU(n) symmetric point, where $S_1 = 0$, we find that
$\Delta_{ij}=1$ for all of the tunneling operators.
This fact is most easily seen by exploiting the fermionic
representation described in section II.
It follows that for $S_1 = 0$
weak disorder is relevant, and
grows stronger under scaling to low energies,
for all $\nu = n/(np+1)$.  Moreover, as discussed in Section IV,
for non-negative $p$
the scaling dimension does not depend on the non-universal
velocities which enter into the action $S_1$.  Thus for all
fillings $\nu$ with non-negative $p$, weak disorder is relevant,
and must be treated non-perturbatively.  For fillings
with negative $p$, such as $\nu=2/3$,
the scaling dimensions $\Delta_{ij}$ will vary with the non-universal
velocities entering in $S_1$.
If these velocities are tuned
so that $\Delta$ exceeds 3/2, then there will be an edge phase
transition into a phase in which disorder is irrelevant.
For filling $\nu = 2/3$ this phase transition was analyzed in Reference
\cite{us23}.
In this paper we will confine our
attention to the phase in which $\Delta_{ij}<3/2$, where the
disorder is relevant.

At finite temperatures, some information can be obtained
using perturbation theory in the impurity strength .  This will
be discussed in section IIIc, and in more detail in appendix
B.  However, it is clear that the
low temperature physics lies outside of the
perturbative regime.  As we now show,
however, it is possible to use the
fermionic representation of the SU(n) symmetric model (with
$S_1=0$) to obtain an exact solution for arbitrary disorder strength.
In section IIIb we go on to show that the resulting random fixed point is
stable to weak perturbations (non-zero $S_1$), so that this soluble
model provides a
description of the low temperature physics in the
entire disorder dominated phase.

\subsection{Exact Solution: The Random Fixed Point}

Consider then the addition of random edge scattering terms (3.3) to
the SU(n) invariant action $S_0 = S_\rho + S_\sigma$ in (2.27).
While such terms naively
break the SU(n) symmetry, the solution below reveals the presence of
a hidden but still exact SU(n) symmetry in this random problem.
It is
useful to first re-express $S_{\rm random}$ in (3.3) in terms
of the fermion fields appearing in $S_0$ in (2.27).  Under the
transformations described in Section IIc,
$\phi_i - \phi_j \rightarrow \tilde\phi_i - \tilde\phi_j$, so
we may identify
\begin{equation}
e^{i(\phi_i - \phi_j)} \rightarrow \psi^\dagger R^{ij} \psi .
\end{equation}
Here $R^{ij}$ is the off diagonal SU(n) generator defined in
section II.
This allows us to re-write $S_{\rm random}$
in terms of fermion fields as
\begin{equation}
S_{\rm random} = \int dx d\tau \psi^{\dagger} M(x) \psi
\end{equation}
where $M(x)$ is a random $n\times n$ matrix,
\begin{equation}
M(x) = \sum_{i > j} [ \xi_{ij} (x) R^{ij} + \xi^*_{ij} (x) R^{ji} ] .
\end{equation}

Notice that the charge sector $S_\rho$ is completely unaffected
by the random tunneling.
In addition, the neutral sector, $S_\sigma + S_{\rm random}$
is purely bi-linear
in the Fermi fields, but with a spatially random 'coefficient',
$M(x)$.  These random terms act as SU(n) symmetry breaking fields
on the quadratic action $S_\sigma$.  However, they can be eliminated
from the action by defining a new set of Fermion fields, $\tilde\psi$,
which are related to the original Fermions via a suitable spatially
dependent SU(n) rotation.  Specifically,
upon defining a new fermion field,
\begin{equation}
\tilde\psi(x) = U(x) \psi(x)
\end{equation}
with a unitary SU(n) rotation
\begin{equation}
U(x) = T_x \exp[{i \over v} \int_{- \infty}^x dx^{\prime} M(x^\prime) ]
\end{equation}
with $T_x$ an x-ordering operator, the action becomes
simply
\begin{equation}
S_0 + S_{\rm random}=
S_{\rho} + \int dx d\tau \tilde\psi^\dagger (\partial_\tau
-iv\partial_x) \tilde\psi  .
\end{equation}

The action is quadratic in terms of these new rotated fermion fields,
with the neutral sector still possessing a full SU(n) symmetry.
We have successfully eliminated all random terms by exploiting the
SU(n) symmetry present in the pure action $S_0$.
Since the transformed action is quadratic and the disorder does
not occur explicitly, we can define a simple RG transformation on $\phi_\rho$
and the rotated fermions, $\tilde\psi$, which
leaves the action invariant.  Our exact solution thus describes
a fixed point, with a U(1) charge symmetry and an SU(n) symmetry
in the neutral sector.  However, it must be borne in mind that we are actually
describing a random fixed point, with correlation functions
of the original fields depending on the randomness via the above
random SU(n) rotation.

\subsection{Stability of Random fixed point}

Having established that the action
$S_0 + S_{\rm random}$ decouples into independent charge and neutral sectors,
we must now take into account the nondiagonal interaction
matrix in $S_1$, which we have ignored above.  Being non-random,
these terms couple the charge and neutral sectors and
break the SU(n) symmetry even after ensemble averaging
over the disorder.  However,
as we now show, these terms are irrelevant
at the random fixed point described by (3.10).
The randomness is crucial to guarantee the irrelevance of these
operators.  As we shall see, without the inclusion
of randomness, which is "hidden" in the representation (3.10),
the symmetry breaking perturbations in $S_1$ are not
driven to zero.

It is convenient to re-express $S_1$ in terms of the fields
appearing in (3.10), namely the charge field $\phi_\rho$,
and neutral fermion fields $\tilde\psi$.
Upon performing the orthogonal transformation described in Section IIc, it is
apparent that $S_1$ in (2.8) can be re-expressed
as a sum of three types of terms:
\begin{equation}
S_{1a} = \int dx d\tau v_a (\partial_x \phi_\rho )^2  ,
\end{equation}
\begin{equation}
S_{1b} = \int dx d\tau \sum_{i,j =1}^{n-1} v^{ij}_b  \partial_x
\Phi_i
\partial_x \Phi_j   ,
\end{equation}
\begin{equation}
S_{1c} = \int dx d\tau \sum_{i=1}^{n-1} v_c^i \partial_x \phi_\rho
\partial_x \Phi_i   .
\end{equation}
%%%%%%%%    In S_{1b} I presume we must include the i=j terms,
%%%%%%%%      but also irrelevant - PLEASE CHECK
The coefficients $v_\mu$ (with $\mu=a,b,c$) can be expressed
in terms of the velocity matrix $V_{ij}$.
The first term above is innocuous, and can be absorbed into
$S_\rho$, giving a shift in the
velocity of the charge mode.
To analyze the other terms it is useful to re-express the
boson fields $\Phi$
in terms of the Fermion fields:
\begin{equation}
{1\over{2\pi}}\partial_x \Phi_m \rightarrow \psi^\dagger
D^m \psi \rightarrow \tilde\psi^\dagger M^m(x)
 \tilde\psi
\end{equation}
where $M^m(x)$ are $n\times n$ matrices given by
\begin{equation}
M^m(x) = U(x)D^mU^\dagger (x)
\end{equation}
with $U(x)$ defined in (3.9).
The unitary matrix $U(x)$ is a random x-dependent SU(n) rotation,
which is uncorrelated on scales long compared to a mean free path
for interchannel scattering, $\ell \sim v_\sigma^2/W$.
Thus $M^m(x)$ will similarly be random $n\times n$ matrices.
Treating $v_b$ as small, we can now show that the SU(n)
fixed point described by (3.10) is stable to this perturbation.
Note first that the operator in $S_{1b}$ involves four
fermion fields, $\tilde\psi$, and so has a scaling dimension
of $\delta =2$, at the SU(n) fixed point described by (3.10).
Since the coefficient of this operator is spatially random,
we consider the linear RG flow equation for its mean
square average, $W_b \propto v_b^2$, which is of the form,
\begin{equation}
{\partial W_b\over{\partial\ell}} = (3-2\delta)W_b.
\end{equation}
The perturbation is clearly irrelevant.
It should be emphasized that in the absence of randomness,
the dimension 2 operators in $S_{1b}$ are marginal and do not
renormalize to zero!  Thus disorder is seen to be absolutely
critical in the stability of the SU(n) fixed point (3.10).
The reason why the random perturbation is irrelevant, while the
uniform perturbation is marginal, can be understood as follows.
The mean square average of
the random perturbation
over a length scale $L >> \ell$,
is an average over $L/\ell$ uncorrelated
regions, and will hence decay as $L^{-1}$.
This accounts for the renormalization group eigenvalue of
$-1$ in (3.16).

The above argument can also be used for
the perturbation $S_{1c}$ in (3.13), which mixes the charge and neutral
sectors.
This operator also has a scaling dimension of $\delta=2$, and
with a spatially random coefficient $v_c$ will likewise scale
to zero under a RG transformation.

We thus see that the disorder has played a crucial role
in both driving the charge/neutral coupling to zero, and
driving the SU(n) symmetry breaking interactions in the neutral sector to zero.
The final fixed point theory, described by (3.10),
has a full U(1)xSU(n) symmetry,
a much higher symmetry than the underlying random Hamiltonian.

\subsection{Finite Temperatures : The Hydrodynamic Regime}

The exact solution (3.10) of the random edge which describes a stable
zero temperature fixed point can also be used to extract
physical properties of the edge at low but non-zero temperatures.  These
properties will
be determined by the structure of the
fixed point itself, and the leading irrelevant operators,
such as those proportional to $v_\mu$ above.  At
low but non-zero temperatures these operators have not had "time"
to fully renormalize to zero, and can then have
an important effect on physical observables.
Although one can show that the irrelevant operators
do not modify the quantized Hall conductance itself,
they do dramatically effect the
propogation of the neutral modes at finite temperature.

To see why, we first note that the existence
of the propagating neutral modes is tied intimately to
the exact SU(n) symmetry in the neutral sector at
the fixed point.  But at finite temperatures, this symmetry
is no longer exact, due to the presence of irrelevant
operators, so that the neutral modes should no
longer be strictly conserved.  Thus, one
expects that at finite temperatures
the neutral modes should decay away at a non-vanishing rate, $1/\tau_\sigma$.
Equivalently, one expects  a finite decay length, or "inelastic
scattering length", $\ell_\sigma= v_\sigma
\tau_\sigma$.  On scales $L$ much larger than
$\ell_\sigma$, the neutral modes should not propagate.
Since the fixed point is approached as $T \rightarrow 0$,
however, the decay length should diverge
in this limit.

At wavelengths long compared to $\ell_\sigma$, we thus expect
a hydrodynamic regime, in which the only propagating
modes are those required by conservation laws.
Since the only conserved quantity in this regime is the total electric
charge, we expect a single propagating "zero sound" mode.

In order to establish the existence of the hydrodynamic
regime and to compute the
temperature dependence of the neutral mode
decay rate we evaluate the self energy of the
neutral mode perturbatively about the random fixed point (3.10).  The dominant
contributions come from the interactions $v_b$ and $v_c$
in equations (3.12) and (3.13).
Notice that
(3.13) contains  terms of the form
\begin{equation}
\delta S = \int dx d\tau \tilde v^{ij}_c(x) \partial_x \phi_\rho
\tilde\psi^\dagger R^{ij} \tilde\psi
+ c.c.
\end{equation}
where $\tilde v^{ij}_c(x)$ is a random coefficient
which depends, as in (3.15), on the random SU(n) rotation.
Breaking the SU(n) symmetry, this term explicitly violates the conservation of
the neutral
modes.

It is convenient at this stage to re-bosonize the
"rotated" fermion fields.  We thus "undo" the steps which
lead us from equations (2.24) to (2.27), writing
$\tilde\psi_i = \exp(i \tilde\chi_i )$, and
$\tilde\chi_i = O_{ji} \chi_j$.  (In the absence of disorder
we would thus have $\chi_i = \Phi_i$ in (2.24)).
In terms of these new bosonic fields the fixed point action in (3.10)
now takes the form
\begin{equation}
S_0 + S_{\rm random} =
S_\rho +  \int dx d\tau {1 \over {4\pi}} \sum_{i=1}^{n-1}
\partial_x \chi_i ( i \partial_\tau + v \partial_x ) \chi_i,
\end{equation}
where we have omitted the auxiliary "charge" mode, $\chi_n$.

In this representation, we can now evaluate the self energy
for $\chi_i$ perturbatively in $\tilde v$.  For simplicity
we consider here only contributions from
the term $\tilde v^{12}$ in (3.17) for which the corresponding operator
has a particularly nice bosonized representation,
\begin{equation}
\partial_x \phi_\rho \tilde\psi^\dagger R^{12} \tilde\psi =
 \partial_x \phi_\rho e^{i \sqrt 2 \chi_1 }  ,
\end{equation}
where
\begin{equation}
\partial_x \chi_1 =
{1 \over {\sqrt 2}} \partial_x (\tilde\chi_1 - \tilde\chi_2 )
=\tilde\psi^\dagger D^1 \tilde\psi.
\end{equation}

Since the perturbation $v_{12}$ only involves $\chi_1$,
its effects will be contained in the retarded Greens function,
\begin{equation}
G^R_1 (x,t) = <[\chi_1(x,t),\chi_1(0,0)]>\theta(t).
\end{equation}
When evaluated at the fixed point (3.18) it takes the simple form,
\begin{equation}
G^{0R}_1 (q,\omega) = {2\pi\over{q(\omega + i\epsilon- v q)}}
\end{equation}
exhibiting a pole at the neutral mode frequency,
$\omega = v q$.

For simplicity we take the random coefficient $\tilde v^{12}$ to be delta
correlated in space, with variance $W_c$.  The self energy
may then be evaluated perturbatively in $W_c$.
To lowest order the
self energy involves the diagrams shown in figure 3.
These are evaluated in appendix B, where we show that
at low frequencies,
\begin{equation}
\Sigma(q,\omega) = {i\omega \over {2\pi\ell_\sigma}},
\end{equation}
with $\ell_\sigma^{-1} \propto W_c T^2$.
Unfortunately, this lowest order approximation to the self energy
leads to an incorrect description of the long wavelength limit,
inconsistent with the hydrodynamic regime.
In Appendix B we show that the correct self energy,
obtained by summing a class of diagrams, is given by
\begin{equation}
\Sigma(q,\omega) = {i\omega\over{2\pi\ell_\sigma}}
{1\over{1 - i (q\ell_\sigma)^{-1}}}.
\end{equation}
Note that to leading order in $W_c$ (or $\ell_\sigma^{-1}$)
(3.24) reduces to (3.23).  However, higher order terms in
the expansion are singular in the $q \rightarrow 0$ limit
and must be accounted for in the correct long wavelength
theory.

Using (3.24) we may then write the
retarded neutral boson Green's function as
\begin{equation}
G^R_1(q,\omega) = 2\pi{1 - i (q\ell)^{-1}
\over{q(\omega - v_\sigma q + i/\tau_\sigma)}}
\end{equation}
which exhibits a neutral mode decaying at a rate
\begin{equation}
{1\over\tau_\sigma} =  v_\sigma/\ell_\sigma \propto  W_c T^2.
\end{equation}
An analogous calculation leads to a similar result for the interactions
given by $v_b$ in equation (3.12).
At non-zero temperatures the neutral mode decays away,
just as for a quasiparticle in a Fermi-liquid.

The effects of
the irrelevant operators on the charge mode may be evaluated
in a similar manner.  However,
due to charge conservation we do not
expect the charge mode to decay.
Indeed, it can be seen explicitly that the interaction terms
in (3.11-3.13) commute with the total charge.
Consider the Greens function for the charge
mode,
\begin{equation}
G^R_\rho(x,t) = <[\phi_\rho (x,t),\phi_\rho (0,0)]>\theta(t)  ,
\end{equation}
which at the fixed point (3.10) is given by
\begin{equation}
G_\rho(q,\omega) = {2\pi\over{q(\omega +i\epsilon - v_\rho q)}}.
\end{equation}
The contribution to the self energy due to the
interaction (3.13) may be computed along the same lines as above,
and we find
\begin{equation}
\Sigma_\rho(q,\omega_n) \propto i\omega q^2 W_c
\end{equation}
This leads to a correction to the charge mode propagator
which becomes
\begin{equation}
G_\rho(q,\omega) = {2\pi\over{q(\omega - v_\rho q + i D q^2)}}.
\end{equation}
This form implies
that a localized charge pulse will spread diffusively as it
propagates down the edge with a temperature independent
diffusion constant $D \propto W_c$.
As expected, though, due to charge conservation the decay rate
vanishes at $q=0$, in contrast to the neutral modes.

By working perturbatively about the random fixed point (3.10),
we have thus shown that at finite temperatures,
on length scales long compared to $\ell_\sigma \propto T^{-2}$,
there exists a hydrodynamic regime characterized by a
single propagating charge mode.  It is also instructive to
recover this hydrodynamic regime by working perturbatively about
the fixed line in the absence of randomness,
described by (2.7-2.8).  This
will be valid at high temperatures when weak disorder has not had "time" to
flow out of the perturbative
regime.  Like the random fixed point, the clean fixed line also
has higher symmetry because each of the $n$ propagating modes are
independently conserved.  As explained in section IIb, this implies
a conductance which is not quantized when channels move
in opposite directions.
However, even weak
interchannel scattering destroys
the independent conservation laws, leaving total charge as the only conserved
quantity.  We thus expect a long wavelength
hydrodynamic regime with only a single propagating charge
mode.  In appendix B we analyze the effects of
randomness perturbatively and establish this hydrodynamic
regime on length scales longer than the mean free path $\ell$ for
interchannel scattering, given by
\begin{equation}
\ell^{-1} \propto W T^{2\Delta-2}
\end{equation}
where $W$ is the r.m.s. strength of the randomness and
$\Delta$ is the scaling dimension of the tunneling operator.
Moreover, we find that in this hydrodynamic regime
the quantization of the conductance is restored,
giving $G = \nu e^2/h$ even when $p<0$.

\section{Tunneling at the Edge}

We now apply the theory described above to compute the scaling dimension
of general edge tunneling operators.  These scaling dimensions determine
the temperature exponents
for tunneling through a
point contact\cite{Moon,us1} between two Hall fluids.

The most general edge tunneling operator can be written
\begin{equation}
\hat T(x) = e^{i\sum_{j=1}^n m_j \phi_j(x)}
\end{equation}
for arbitrary integers $m_j$.
This operator creates an edge quasiparticle excitation at
position $x$, with charge $Q$ given by
\begin{equation}
Q= \sum_{ij} m_i K^{-1}_{ij} .
\end{equation}
This is the same value as the bulk quasiparticle charge (2.5).
For filling factor $\nu=n/(np+1)$, the inverse of the
K-matrix is, $K^{-1}_{ij} = \delta_{ij}
- p/(np+1)$, which gives for the charge,
\begin{equation}
Q=\nu \bar{m}  ,
\end{equation}
with the definition
\begin{equation}
\bar{m} = {1 \over n} \sum_{j=1}^n m_j  .
\end{equation}

For non-negative p the charge and neutral modes
propagate in the same direction.  In this case, we prove in appendix
A that even in the absence of randomness,
the scaling dimension of the tunneling operator is independent
of the (non-universal) velocity matrix in $S_1$ (2.8).
This can be understood by
considering the form of the correlation function computed in the
absence of randomness,
\begin{equation}
P(x,\tau) = <\hat T(x,\tau)^\dagger \hat T(0,0)> \propto \prod_{i=1}^n
{1\over{ (v_i\tau + ix)^{\delta_i}}},
\end{equation}
where the expectation value is taken with respect to the action $S_0+S_1$
in (2.7-2.8).
Each eigenmode of the quadratic action contributes one term to
the product in (4.5).
The scaling dimension of $\hat T$
is then given by
$2\Delta = \sum_i \delta_i$.  There is a constraint on the
$\delta_i$, though, due to the statistics of
the quasiparticle.  The operator $\hat T$ creates an edge quasiparticle
which must have the same statistics as a bulk quasiparticle.
The statistics of the edge quasiparticle can be defined as
the phase accumulated upon
rotating $x,\tau$ to $-x,-\tau$ clockwise in the Euclidean plane,
since for
$\tau=0$ this effectively interchanges two quasiparticles.
In appendix A we verify by explicit calculation that the "edge"
statistics angle defined in this way indeed
equals the bulk statistics angle given in (2.6).
Since for $p \ge 0$ all of the velocities $v_i$ have the same
sign, under this exchange $P \rightarrow P \exp i\Theta$, with a
statistics angle $\Theta/\pi = \sum_i \delta_i$.
Generally, the statistics angle
is a topological property of the bulk quantum Hall state,
is universal and independent of the edge interaction
matrix $V_{ij}$.  But for non-negative $p$ the scaling
dimension of the edge tunneling operator equals the statistics
angle, $2\Delta = \Theta/\pi$, and is therefore also universal.

To describe tunneling at a point contact, the relevant quantity is
the "local" scaling dimension of the edge tunneling operator,
defined via $P(x=0,\tau) \sim \tau^{-2\Delta}$.
This average is independent of the spatially random SU(n) rotation
of Section III.  Thus for $p \ge 0$,  the "local" scaling dimension
is still given by the (bulk) quasiparticle statistics, even in the presence
of disorder.  We thus have from (2.6),
\begin{equation}
2\Delta = \sum_{ij} m_i K^{-1}_{ij} m_j.
\end{equation}
For bulk filling $\nu=n/(np+1)$
this can be written as
\begin{equation}
2\Delta = {Q^2 \over \nu} + \sum_{j=1}^n (m_j^2 - \bar{m}^2 ) .
\end{equation}

When p is negative the charge mode propagates in the direction
opposite to the neutral modes, so that the velocities in (4.5) are
no longer all of the same sign.  The constraint imposed by the
bulk quasiparticle statistics therefore becomes
$\Theta/\pi = \sum_i {\rm sgn} (v_i) \delta_i$, and no longer
determines the scaling dimension of the edge tunneling operator, $2\Delta
= \sum_i \delta_i$.  Thus,
in the absence of random tunneling terms,
$S_{\rm random} =0$, the scaling dimension of $\hat T$
is non-universal depending on the
velocity matrix $V_{ij}$ in $S_1$.  However, with randomness present,
the system flows to the fixed point (3.10), which has an exact U(1)xSU(n)
symmetry.  The "local" scaling dimension
of the edge tunneling operators, $\hat T$, then follow from
the universal properties of this fixed point.

To evaluate them it is useful to re-express the tunneling operator
in terms of the charge and neutral fields.
We find,
\begin{equation}
\hat T =  e^{i\bar{m} \phi_\rho} \prod_{i=1}^n \psi_i^{(m_i - \bar{m})}  .
\end{equation}
Being interested only in the local scaling dimension we can replace
$\psi$ by $\tilde\psi$ in the above.  Then upon evaluating
the average in (4.5) using the quadratic fixed point action (3.10) gives
\begin{equation}
2\Delta = {Q^2 \over |\nu|} + \sum_{j=1}^n (m_j^2 - \bar{m}^2 )
\end{equation}
The first term comes from the charge mode, and the second
contribution from the neutral sector.  Notice that this form
is the same as that for $p \ge 0$, except with a modulus of $\nu$.

We note here that the irrelevant operators discussed in
section IIIC, which lead to a finite lifetime for the neutral mode at
finite temperatures,
do not affect our results for the tunneling exponents.
The finite lifetime introduces a $1/\tau_\sigma \propto T^2$
cutoff into the logarithmically
divergent integral which occurs in the exponent when
the correlation function $ <\hat T(\tau) \hat T(0)> $ is computed.
But this divergence is already cut off by the temperature, $T$,
so at low temperatures the finite lifetime has no effect.
Thus even though the neutral mode is not conserved at finite
temperature, it has a crucial effect on the asymptotic temperature
dependence of the tunneling exponents.

With our final expressions for the "local" scaling dimensions of
the most general edge tunneling operators, we are in position to
make quantitative predictions for a number
of intersting experiments.

\section{Experimental implications}

The simplest experiment which is sensitive to the
edge dynamics involves making a constriction or point contact
in a quantum Hall fluid.  At the constriction,
the top and bottom edges of the Hall bar are close together,
as shown in figure 4,
facilitating tunneling processes between the two edges.
Any charge which tunnels between the edges is
effectively backscattered, and will reduce the source
to drain conductance.  The "local" scaling
dimension of the edge tunneling operators will then feed into
the temperature dependence of the conductance through the constriction.

Consider first the limit of a very slight constriction,
which will give a small amount of backscattering.
In this regime, the temperature dependence of the conductance through
the constriction will be dominated by the quasiparticle
which can tunnel most easily from top to bottom edge.  The amplitude for such
tunneling, though, will in general be temperature dependent,
and at low temperatures will be dominated by
the quasiparticle with the smallest scaling
dimension.  To leading order the conductance will take the form\cite{us1}
\begin{equation}
G(T) = |\nu| {e^2\over h} - v^2 T^{2(2\Delta_{\rm min}-1)}+O(v^4) ,
\end{equation}
where $v$ is the amplitude for the tunneling of
this quasiparticle.

A charge $Q=\nu$ quasiparticle with scaling dimension
$2\Delta = |\nu|$ can be obtained from (4.1) by taking
$m_j=1$ for all $j$.
On the other hand, a single $m_j=1$ with the rest equal
to zero creates an n fold degenerate excitation with charge $\nu/n$ and
scaling dimension
$2\Delta = 1 - 1/n - |\nu|/n^2$.  For the integer quantum Hall effect,
$\nu=n$ ($p=0$), the latter corresponds to a quasiparticle
with the smallest dimension, $2\Delta_{min} =1$.  Physically, it corresponds
to tunneling an electron into one of the $n$ edge channels.
As expected, the backscattering which reduces the conductance in
(5.1), is independent of temperature in this case.

For all fractional quantum Hall states ($p \ne 0$) {\it except}
$\nu = 2/3$ ($n=2$,$p=-2$), the charge $\nu$ quasiparticle has the
smallest dimension, so that
\begin{equation}
2\Delta_{\rm min} = |\nu|.
\end{equation}
Notice that with $|\nu| < 1$ the backscattering corrections
grow at low temperatures.  The above form (5.1) is only valid
down to temperatures where the corrections to the
quantized conductance remain small.

By varying parameters, such as a gate voltage,
it should be possible to tune the amplitude of the
leading relevant backscattering to zero.  This will appear as a resonance
in the conductance.
The robustness of the resonances at low
temperatures, that is the temperature dependence of the resonance
peak, will be determined by
the backscattering of the quasiparticle with the next smallest
scaling dimension.  If this scaling dimension
is larger than 1/2 then the
conductance on resonance will be perfect, and the resonance robust.
In this case one expects that the resonance lineshape
should be universal, as predicted for $\nu =1/3$.
%Are there any cases for which this is true?????????
If less than 1/2, the resonance peak will diminish in amplitude
upon cooling
and eventually vanish completely in the
zero temperature limit.  In either case,
there should be a regime in temperature where the width of the resonance
narrows upon cooling, varying as $T^{(1-|\nu|)}$.

For the special case $\nu=2/3$ it turns
out that the quasiparticle tunneling operators with $(m_1,m_2) = (1,0),(0,1)$
and $(1,1)$ all
have the same dimension.  The dimension is still given by
(5.2), but now the most relevant operator
is not unique, but three fold degenerate.
This suggests
that resonances for $\nu=2/3$ will tend to be
less robust than for the other fractions.

Away from resonances, in the fractional Hall effect ($p \ne 0$),
the conductance through the constriction drops with decreasing
temperature, eventually invalidating the perturbative result (5.1).
In this regime, we can consider the opposite limit of
large backscattering, where the dominant process is
physical
electrons tunneling through the point contact from one side to the other.
In this limit, the conductance will be dominated
by the charge $Q=1$ tunneling operator, which has the smallest
scaling dimension, varying with temperature as\cite{us1}
\begin{equation}
G(T) \approx t^2 T^{2(2\Delta_{\rm min}(Q=1)-1)}  .
\end{equation}
Here $t$ is the amplitude for the dominant electron tunneling process.
The scaling dimension $\Delta_{\rm min}$ for $Q=1$ can be obtained by taking
$m_j = p + \delta_{jk}$ for $k=1,...,n$,  in (4.1), which gives
\begin{equation}
2 \Delta_{\rm min} (Q=1) = {1 \over |\nu|} + 1 - {1 \over n} .
\end{equation}
This tunneling operator is not unique, but n-fold degenerate.

For non-negative $p$, when all the edge channels are moving in the
same direction,
the above two expressions can be combined to give
\begin{equation}
G(T) \approx t^2 T^{2p} .
\end{equation}
For the integer quantum Hall effect this gives the expected
temperature independent result.
For the dominant $p=2$ sequence, $\nu=1/3,2/5,3/7,4/9,...$,
this predicts a $T^4$ temperature dependence.

When $p$ is negative, and the neutral modes are moving in a direction
opposite to the charge modes,
the conductance can be expressed using (5.3-5.4) as
\begin{equation}
G(T) \approx t^2 T^{2|p|-(4/n)}  .
\end{equation}
For the $p=-2$ sequence, $\nu=2/3,3/5,4/7,5/9,...$
the predicted power laws are $2, 8/3, 3, 3 1/5,...$  which approach
a $T^4$ as $\nu$ approaches 1/2.  It is worth emphasizing that in this case
the particular powers are determined by the structure of
the disorder dominated fixed point, at which the neutral and
charge sectors decouple.  Even though it is electron tunneling
which dominates in this regime, the neutral modes are essential.
Since an
electron is built from a superposition
of the charge mode and neutral modes, upon tunneling through
the point contact into the edge, the electron excites both the charge
and neutral modes.
%%%   SHOULD WE HAVE A TABLE HERE?????????  Yes.

Another central feature of
the random U(1)xSU(n) fixed point, is that
the $n-1$ neutral modes are all predicted
to move at the same velocity!  So, for example, at the
edge of an integer quantum Hall state with filling $\nu=3$,
the three edge modes, which will in general have different
velocities in the absence of edge randomness, are predicted
to de-couple with randomness into a charge mode, moving
at one velocity, and two neutral modes moving at the same velocity
as one another.  This de-coupling will take place on length scales
longer than an edge mean free path.  The mean free path
depends on the strength of the edge impurity scattering and the spatial
separation between the various edge modes, and will thus clearly
be a sample specific length.  For $\nu=2$ one has a rather nice
example of "spin-charge" separation, with the edge
channel index playing the role of the electron spin $s_z$.  The disorder
decouples the charge mode from the SU(2) invariant neutral mode,
the analog of the "spin mode", and the two modes
separate, moving at different velocities.

The edge neutral modes might be directly measurable via suitable
low temperature time
domain transport experiments, similar to Ashoori et. al.\cite{Ashoori}.  In
Ashoori et. al. the edge states of a quantum Hall sample were excited
by sending a short pulse into a capacitor placed near the edge.
Another capacitor was used to detect
the propagating edge modes, on the other side of the sample.
For filling $\nu =2/3$ only one propagating mode was observed.
Our theory predicts the existence of only one charge mode,
the other being neutral and presumably coupling very weakly to
the capacitors.  This could explain
naturally the observed absence of a second propagating mode.
A suitable generalization of Ashoori et. al. which
would allow for detection of the neutral modes would be
to replace the capacitors with tunnel junctions.
Sending a short pulse of electrons into the edge of a
quantum Hall sample, would excite both the charge and neutral modes
at the edge.  Provided the temperature was low enough that
the decay length $l_\sigma$ exceeds the sample
dimensions,
the neutral modes could be
detected at the far side of the sample with another tunnel
junction.  The neutral
modes, upon passing by the second tunnel junction, would excite
electrons to tunnel into the leads, and should be detectable
as a time domain current pulse.  By varying the temperature
it might also be possible to extract the temperature dependence
of the neutral mode decay rate, to test the predicted
$T^2$ dependence.  This would be the analog of a direct real time
measurement of a decaying Fermi-liquid quasiparticle!

In this article we have established the
existence of a random edge fixed point for states
at filling $\nu =n/(np+1)$ (with U(1)xSU(n) symmetry),
and demonstrated that it is locally stable.  It should be
emphasized, however, that we have not argued for the absence
of other edge phases, at the same bulk filling.  It is conceivable
that for a given filling $\nu$, there exist other edge phases, separated from
the phase we have analyzed by an edge phase transition.
In fact, for the special case of $\nu=2/3$ we know this to be the case.
In our earlier paper with Joe Polchinski we found another (locally)
stable edge fixed point
for $\nu=2/3$, at which weak
random edge tunneling was irrelevant and the charge and neutral sectors
did not de-couple.  In this phase
the two terminal conductance at
$T=0$ is non-universal.  However, at finite temperatures the
quantization of the conductance is restored, provided
the sample is larger than the mean free path for interchannel tunneling
(which diverges as $T\rightarrow 0$).
Moreover, in this disorder-free phase
the tunneling exponents were predicted to be non-universal.
There was a Kosterlitz-Thouless like zero temperature
phase transition separating the two phases.
However, for filling $\nu$ with non-negative $p$,
the disorder free edge phase is always perturbatively
unstable to disorder.
Thus if other edge phases exist for these fillings,
they will presumably also
be described by random fixed points.  Ultimately though,
the actual phase for the edge of a given real
sample will have to be determined by comparing with the predicted behavior.

It is also worth pointing out that the special SU(n) symmetry
at the edge has only been established for the class of
Hall states at filling
$\nu = n/(np+1)$, with $n$ integer and $p$
an even integer.  For other filling fractions not of this
form, such as $\nu=4/5$, there may not be such high symmetry at the edge.
The low energy edge structure at these fillings will be
the subject of future work.

In brief summary, we have shown that disorder at the edge of a quantum Hall
fluid plays an essential role
in determining the structure of the low energy edge excitations.  In
particular, for fractional quantum Hall states at filling
$\nu = n/(np+1)$, we have shown that the disordered edge actually has a
higher symmetry than a perfectly clean edge would have.  The charge
is carried in a single mode, and the remaining $n-1$ neutral modes
all propagate at the same speed and possess an SU(n) symmetry.
An exact solution for the random SU(n) fixed point has been presented,
which allows for numerous quantitative experimental predictions.

This paper is an extension and generalization of
an earlier paper with Joe Polchinski (Ref. \onlinecite{us23}).
We are extremely grateful and indebted to Joe for many of
the key ideas and results contained herein.
We are also grateful to N. Read, R. Webb and A. Zee
for fruitful discussions.  M.P.A.F. has been supported by the
National Science Foundation
under grants No. PHY89-04035 and No. DMR-9400142.

\appendix
\section{General Treatment of Multichannel Clean Edge}

In this appendix we consider a clean multichannel edge described
by the general action
\begin{equation}
S = \int dx d\tau {1\over {4\pi}} \sum_{ij}
\partial_x\phi_i  \left[  K_{ij}\ i \partial_\tau  +
v_{ij} \partial_x
\right] \phi_j.
\end{equation}
We wish to compute the conductance as well as the scaling dimension
and statistics angle of
tunneling operators.  We show that these quantities are universal
if all of the channels propagate in the same direction.  In
general, however, the conductance and scaling dimensions are
non universal.  The following analysis is valid for an arbitrary
$K$ matrix, and is not limited to quantum Hall edge states at filling factors
$\nu = n/(np+1)$.

In order to proceed, it is convenient to transform
the problem into a representation in which both $K_{ij}$ and
$v_{ij}$ are diagonal.
This can be accomplished in
three steps.  First we diagonalize the matrix $K$ via an orthogonal
transformation
$\phi_i = \Lambda_{1,ij} \tilde\phi_{1,j}$, where
$(\Lambda_1^T  \Lambda_1)_{ij} = \delta_{ij}$ and
$(\Lambda_1^T K \Lambda_1)_{ij} = \lambda_i\delta_{ij}$.
For $\nu = n/(np+1)$
this transformation was performed explicitly in section
II:  $\Lambda_{1ij} = O_{ij}$ defined in (2.21) and (2.22),
and $\lambda_i = 1 + np \delta_{in}$.
We then rescale $\tilde\phi_{1,i}$ by writing
$\tilde\phi_{1,i} = \Lambda_{2,ij} \tilde\phi_{2,j}$, where
$\Lambda_{2,ij} = \delta_{ij}/\sqrt{|\lambda_i|} $.
In this representation the action is
\begin{equation}
S = \int dx d\tau {1\over {4\pi}} \sum_{ij}
\partial_x\tilde\phi_{2i}  \left(  \eta_{ij} \ i\partial_\tau  +
\tilde v_{ij} \partial_x \right) \tilde\phi_{2j}
\end{equation}
with $\tilde v= \Lambda_2^T \Lambda_1^T (v_{ij})
\Lambda_1 \Lambda_2$,
and $\eta_{ij} = {\rm sgn}(\lambda_i)\delta_{ij}$.
It is now possible to diagonalize $\tilde v$ via a transformation
which preserves the form of $\eta_{ij} $.
Thus, we let $\tilde \phi_{2,i} = \Lambda_{3,ij}\tilde\phi_{3,j}$
where, $(\Lambda_3^T \tilde v \Lambda_3)_{ij} = \tilde v_i \delta_{ij}$
and $\Lambda_3^T \eta \Lambda_3 = \eta$.  $\Lambda_3$ will have the
form of a Lorentz transformation in which ${\rm sgn}(\lambda_i) = +1$
and $-1$ correspond to spacelike and timelike dimensions.
We have thereby decoupled the channels, so that
\begin{equation}
S = \int dx d\tau {1\over{4\pi}} \sum_i
\partial_x\tilde\phi_{3i} \left(
 \eta_i \ i \partial_\tau + \tilde v_i \partial_x
\right) \tilde\phi_{3i}
\end{equation}
This gives an explicit description of the eigenmodes of the system.
Since stability of the action requires that $\tilde v_i >0$,
the direction of propagation of each mode is determined by
$\eta_i$  ($=\eta_{ii}$).

It is now straightforward, using the technique outlined
in section IIB, to formally compute
the edge current,
\begin{equation}
I = \sum_{ij}  (\Lambda_1 \Lambda_2 \Lambda_3)_{ij}
\dot{\tilde\phi_{3j}}/2\pi  ,
\end{equation}
in response to the applied potential $V(x)$,
which couples to the total charge density
\begin{equation}
\rho(x) = \sum_{ij}  (\Lambda_1 \Lambda_2 \Lambda_3)_{ij}
\partial_x{\tilde\phi_{3j}}/2\pi  .
\end{equation}
If different channels
move in opposite directions, then the current will depend on the
voltages in both the left and right reservoirs, $V_L$ and $V_R$.
We find
\begin{equation}
I_{\rm edge} = {e^2\over h} \sum_{ij} (M^+_{ij} V_L - M^-_{ij} V_R)
\end{equation}
where
\begin{equation}
M^\pm = \Lambda_1 \Lambda_2 \Lambda_3 {1 \pm \eta\over 2}
\Lambda_3^T \Lambda_2^T \Lambda_1^T.
\end{equation}
We may thus write
\begin{equation}
I_{\rm edge} = {e^2\over h} \left( g_+ V_L - g_- V_R \right) ,
\end{equation}
with $g_\pm$ being dimensionless right/left conductances:
\begin{equation}
g_\pm = \sum_{ij}   M_{ij}^\pm .
\end{equation}
Noting that $\Lambda_3 \eta \Lambda_3^T = \eta$, it is straightforward
to show that
\begin{equation}
M^+ - M^- =  K^{-1} ,
\end{equation}
which is universal and independent of the velocities $v_{ij}$.
It then follows from equations (A9) and (2.3) that
\begin{equation}
g_+ - g_- = \nu.
\end{equation}
When all of the channels move in the same direction, say $\eta =1$,
then $M_-$ is equal to zero.  In this case,
$M^+ + M^-$ (and hence $g_+ + g_-$) is also universal.
However, when there are channels moving in opposite directions,
no such simple relation exists for $M^+ + M^-$.
For $\nu = n/(np+1)$ it is possible to use the explicit form
of $\Lambda_1$ and $\Lambda_2$ to write
\begin{equation}
g_+ + g_- = \nu (\Lambda_3\Lambda_3^T)_{nn}.
\end{equation}
In general, $\Lambda_3$ will depend on the non universal
parameters $v_{ij}$.
It should be noted, however, that when $v_{ij}$ is diagonal, as is
the case at the SU(n) symmetric fixed point, then
$\tilde v_{ij}$ is also diagonal, so that $\Lambda_3=1$.
In this case $g_+ + g_- = \nu$ even when there are channels
moving in opposite directions.

The scaling dimension of a general tunneling operator
\begin{equation}
\hat T = e^{i\sum_i m_i \phi_i}
\end{equation}
may be deduced from the correlation function,
\begin{equation}
P(x,\tau) = < \hat T(x,\tau) \hat T(0,0) >.
\end{equation}
Using the transformations defined above, this may be simply
computed and at zero temperature has the form,
\begin{equation}
P(x,\tau) \propto \prod_k {1\over{(\eta_k\tilde v_k \tau + i x)^{\delta_k}}}
\end{equation}
where the exponent
\begin{equation}
\delta_k = \sum_{ij} m_i (\Lambda_1 \Lambda_2 \Lambda_3)_{ik}
   (\Lambda_3^T \Lambda_2^T \Lambda_1^T)_{kj} m_j
\end{equation}
The scaling dimension is then determined by
\begin{equation}
2\Delta = \sum_k \delta_k = \sum_{ij} m_i (M^+_{ij} + M^-_{ij}) m_j.
\end{equation}
The "edge" statistics angle, as discussed in section IV, is given by
\begin{equation}
{\Theta_{\rm edge}\over\pi} =
\sum_k \eta_k \delta_k = \sum_{ij} m_i K_{ij}^{-1} m_j,
\end{equation}
where we have used (A10).
Thus, we see that $\Theta_{\rm edge}/\pi$ is universal and is equal to
the bulk quasiparticle statistics angle (2.6).  However,
the scaling dimension $2\Delta$ is
only universal when all of the channels move in the same direction.
Otherwise, it is non universal and depends on $v_{ij}$.

The neutral tunneling operators, which correspond to the tunneling
of charge between channels on a given edge, are a special case of the general
tunneling operator described above.
For $\nu = n/(np+1)$ an arbitrary neutral operator may be
written as
\begin{equation}
\hat T_{Q=0} = \exp i\sum_{i=1}^{n-1} n_i \Phi_i,
\end{equation}
where $\Phi_i$ is defined in (2.20).  The scaling dimension
of this operator may then be shown to be
\begin{equation}
2\Delta_{ij} =  \sum_{ij=1}^{n-1} n_i (\Lambda_3\Lambda_3^T)_{ij} n_j
\end{equation}

\section{Edge Dynamics at Finite Temperatures}

In this appendix we analyze the long length scale
edge state dynamics at finite temperatures.
Such an analysis arises in two different contexts.  In
section III, we described the disorder dominated $T=0$ fixed
point which has an SU(n) symmetry, and hence $n-1$ propagating
neutral modes in addition to the charge mode.
In this case, the leading irrelevant operator
which couples the neutral and charged sectors, destroys the
SU(n) symmetry and hence violates the conservation of
the neutral modes.  Thus at finite temperature, when such
operators have not flowed to zero, we expect that the
$n-1$ neutral modes will not propagate on long length
scales.   In this hydrodynamic regime there
should be only a single
propagating mode associated with the conserved electric charge.

An analogous situation arises in perturbation theory
in the impurity scattering strength about
the clean fixed point, described by $S_0 + S_1$ in (2.7-2.8).  In this case,
however,
the perturbation theory is generally divergent at zero temperature.
Nonetheless, at finite temperatures, perturbation theory
can provide some useful information.
Like the random fixed point, the clean fixed point has
a high symmetry,
since the charges in each of the $n$
channels are independently conserved, leading to
$n$ propagating modes.  As shown in appendix $A$, this implies
a nonuniversal conductance when any of the channels move in
opposite directions.  At finite temperatures, however,
the interchannel impurity scattering will destroy
the independence of the different channels.  We thus again expect
a long wavelength
hydrodynamic regime in which only a single propagating charge
mode should exist.  Moreover, as we shall show below, in this
hydrodynamic regime, the conductance is universal and given
by $G= \nu e^2/h$.

In this appendix, we wish to explicitly compute the
Green's functions for the
edge modes in the hydrodynamic regimes described above.
In doing so, we shall obtain the temperature
dependence for the decay lengths for the neutral modes.
The simplest approach
is to develop an approximation for the self energy of the
edge modes.
However, we find lowest order perturbation theory for the
self energy fails to describe the long wavelength limit
of the edge dynamics correctly.  Below we will
explain the origin of this failure and physically motivate
a more accurate description.

Because it is conceptually simpler, we will first focus
on the effects of weak impurity scattering in the vicinity
of the clean fixed point.  The following discussion
can easily be generalized to describe the corresponding physics
in the disorder dominated phase.

For simplicity we
will consider the specific case of $\nu=2/(2p+1)$
in which there are only two modes.
The generalization to other hierarchical quantum Hall states
is straightforward.
In section IIC we showed that these two modes may be described in
terms of an
edge charge density $\partial_x\phi_\rho$
and a single neutral
density, $\partial_x\phi_\sigma$ (where we have defined
$\phi_\sigma = \Phi_1$).
The total action can be written,
\begin{equation}
S = S_0 + S_1 + S_{\rm random } +
\int dx d\tau  ( \eta_\rho\phi_\rho + \eta_\sigma\phi_\sigma)
\end{equation}
Here
\begin{eqnarray}
\nonumber
S_0 + S_1 = && \int dx d\tau
{1\over{4\pi}}[
{1\over\nu}\partial_x\phi_\rho( i\partial_\tau + v_\rho\partial_x) \phi_\rho \\
&& - \partial_x\phi_\sigma(  i\partial_\tau - v_\sigma\partial_x) \phi_\sigma
  + 2 v_{\rm int}\partial_x\phi_\rho \partial_x\phi_\sigma  ]
\end{eqnarray}
describes the clean edge and
\begin{equation}
S_{\rm random} = \int dx d\tau (\xi(x) e^{i\sqrt{2}\phi_\sigma} + c.c.)
\end{equation}
is a weak perturbation which describes random impurity scattering between the
two
channels.  As usual we take
$\xi(x)$ to be $\delta$-correlated with variance $W$.  $\eta_\rho$ and
$\eta_\sigma$ are source terms which may
be used to generate Green's functions.

In the absence of interchannel tunneling, the retarded
Green's functions
\begin{equation}
G_{ab}^{0R}(x,t) =
< [\phi_a(x,\tau),\phi_b(0,0)]>\theta(t),
\end{equation}
with $a,b=\rho,\sigma$,
may be determined from (B2) by analytic continuation
$i\omega_n\rightarrow \omega$,
\begin{equation}
G^{0R}_{ab}(q,\omega_n) =  2\pi  \left(
\begin{array}{cc}
{1\over\nu}q(\omega - v_\rho q)  & v_{\rm int} q^2 \\
v_{\rm int} q^2 & - q( \omega + v_\sigma q)
\end{array}
\right)^{-1} .
\end{equation}
To analyze the effects of the random tunneling,
we begin by evaluating the self energy to
leading order in $W$.
Since $S_{\rm random}$ only involves $\phi_\sigma$,
the only nonzero element of the self energy matrix is $\Sigma_{\sigma\sigma}$.
Evaluating
the diagrams shown in figure 5 gives
\begin{equation}
\Sigma^{(1)}_{\sigma\sigma}(q,\omega_n) =
W \int_0^\beta d\tau (e^{i\omega_n\tau} - 1) P(\tau) ,
\end{equation}
where
\begin{equation}
P(\tau) = < T[e^{i\sqrt{2}\phi_\sigma(\tau)} e^{-i\sqrt{2}\phi_\sigma(0)}]>
 \propto \left[ {\pi/\beta} \over \sin (\pi\tau/\beta) \right]^{2\Delta}.
\end{equation}
Here $\Delta$ is the
scaling dimension of $\exp i\sqrt{2} \phi_\sigma$.
Equations (B6) and (B7) are identical to the formula for the
current current correlation function of a point contact connecting
two Luttinger liquids, which is related to the conductance of
the point contact.  We may thus use the results of reference \cite{us1}
to analytically continue to real time and evaluate the retarded
self energy,
\begin{equation}
\Sigma^{(1)R}_{\sigma\sigma}(q,\omega) = {i\omega \over {2\pi\ell}} ,
\end{equation}
where
\begin{equation}
\ell^{-1} \propto W T^{2\Delta-2}.
\end{equation}
Below we will interpret $\ell$ as a mean free path for interchannel
scattering.
Unfortunately, this leading order
approximation to the self energy does not correctly describe
the long wavelength physics of an edge.
In particular, it predicts the existence of low frequency
modes which are inconsistent with the hydrodynamic arguments
presented above.

In order to understand the origin of this failure and to
physically motivate a way to correct it, it is useful
to analyze the Heisenberg equations of motion satisfied
by the operators $\phi_\rho$ and $\phi_\sigma$.
These
may be derived from (B2) and (B3), and take the form
\begin{equation}
{1\over{2\pi}}\left[{1\over\nu}(\partial_t + v_\rho\partial_x)
\partial_x\phi_\rho
+ v_{\rm int} \partial_x^2\phi_\sigma \right] = \eta_\rho
\end{equation}
\begin{equation}
{1\over{2\pi}}\left[(- \partial_t + v_\sigma\partial_x) \partial_x\phi_\sigma
+ v_{\rm int} \partial_x^2\phi_\rho\right] =  I_\perp + \eta_\sigma,
\end{equation}
where $I_\perp$ is an interchannel tunneling operator given by
\begin{equation}
I_\perp = -i ( \xi(x) e^{i\sqrt{2}\phi_\sigma} - c.c.).
\end{equation}
In the absence of interchannel tunneling the propagation of
the densities $\partial_x\phi_\rho$ and
$\partial_x\phi_\sigma$ is described by the
left hand side of (B10) and (B11).  In this case, there will be two
eigenmodes, which move at different velocities and are in
general linear combinations of $\phi_\rho$ and $\phi_\sigma$.

The $I_\perp$ in (B12) describes the effect of interchannel
tunneling.  During a tunneling event, a unit of charge is
transferred between the channels.
{}From (B10) we see that this
has no immediate effect on the total electric charge density,
$\partial_x\phi_\rho$.  However, (B11) shows that
a well localized
spike of integrated weight $2\pi$ is added to the neutral density
$\partial_x\phi_\sigma$, which in effect measures the charge difference
between the two channels.  Equivalently, a soliton is created,
in which $\phi_\sigma$ winds by $2\pi$.

A linear approximation for the equations of motion may be derived
from the self energy for $\phi_\sigma$ in (B8).
In particular,
Dyson's equation $G^{-1} = G_0^{-1} - \Sigma$ with $\Sigma$ in (B8)
is equivalent to equations (B10) and (B11), with
$I_\perp$ replaced by
\begin{equation}
I_\perp^{(1)} = {1\over \ell} \dot\phi_\sigma.
\end{equation}
This approximation makes physical sense if we
identify  $\dot\phi_\sigma(x)$, via a Josephson like relation,
with the voltage difference between the two channels at point $x$.
Then (B13) is simply a statement of Ohm's law for the tunneling
current at $x$.

However, there is a subtle problem with the interpretation of
$\dot\phi_\sigma$ as a voltage drop which can be seen
from (B10) and (B11).  Let us suppose that at $x_1$,
far away from $x=0$,
an electron tunnels between the two channels at time $t=0$.  Then,
according to (B10), there is a delta function "glitch"
in $\dot\phi_\sigma(x=0)$ at time $t=0$.  This occurs
because the tunneling event introduces a soliton in $\phi_\sigma$ at
$x=x_1$, which forces $\phi_\sigma(x=0)$ to jump by
$2\pi$.   But this "glitch" in $\dot\phi_\sigma(x=0)$ cannot
correspond to a voltage glitch at $x=0$ since it must take a finite
time for the signal
to propagate there.  The origin of this discrepancy is the fact
that $\phi_\sigma$ is an {\it angular} variable, so that
a sudden jump by $2\pi$ should have no effect at all.
There will be a voltage glitch at
$x=0$ only when the soliton in $\phi_\sigma$
(which has a small but finite spatial extent set by the cutoff)
propagates through $x=0$.

Clearly, inclusion of $2\pi$ glitches in
$\dot\phi_\sigma$ in the expression for $I_\perp^{(1)}$
is not physically correct, and is an artifact of the lowest order
approximation for the self energy.
We may correct this situation by replacing equation (B13) with
a modified approximation for the tunneling current, which has the
same physical content in terms of Ohm's law, but does not
include the $2\pi$ glitches in the voltage.  We thus substitute
$\dot\phi_\sigma$ from (B11) into (B13), but then
explicitly remove the term from (B11) involving $I_\perp$, which
only gives the $2\pi$ glitches.
We thereby obtain
\begin{equation}
I_\perp^{(2)} = {1\over {2\pi\ell}} \left( v_\sigma\partial_x \phi_\sigma +
v_{\rm int} \partial_x \phi_\rho  + {2\pi\over\partial_x} \eta_\sigma
\right)  .
\end{equation}

Using the equations of motion with this modified tunneling, including
the source terms, we can
derive a new approximation for the Green's functions,
\begin{equation}
G_{ab}(q,\omega) = 2\pi  \left(
\begin{array}{cc}
{1\over\nu}q(\omega - v_\rho q) & v_{\rm int} q^2 \\
v_{\rm int} q^2 & -q ( {\omega \over{1 - i(q\ell)^{-1}}} + v_\sigma q)
\end{array}
\right)^{-1}   .
\end{equation}
This corresponds to a neutral mode
self energy
\begin{equation}
\Sigma^{(2)}_{\sigma\sigma}(q,\omega) =  {i\omega \over {2\pi \ell}}
{1 \over{1 - i(q\ell)^{-1} }}.
\end{equation}
To leading order in $W$ (or $\ell^{-1}$)
$\Sigma^{(1)}_{\sigma\sigma}$ and
$\Sigma^{(2)}_{\sigma\sigma}$ are equivalent.  However,
(B8) breaks down when $q\ell < 1$.  In order to describe
correctly the long wavelength limit, it is essential to use
the modified self energy (B16).

The validity of this approximation, which we have motivated physically,
may be checked in two ways.  First, it is clear from (B16) that
terms in the self energy at higher orders in $W$ are singular
in the $q\rightarrow 0$ limit.
We may verify this explicitly by considering the self energy
to order $W^2$.  This involves the expectation value of a
product of four of the tunneling operators in (B3).  In evaluating
this self energy, care must be
taken to subtract off the terms in the expectation value
which are one particle reducible, and hence already
accounted for by the leading order term in $\Sigma$.
The resulting term contains precisely the required
singularity, $\omega / ( 2\pi \ell^2 q ) $.  Evidently,
the approximation (B16) corresponds to summing a class
of diagrams which corresponds to a geometric series in
$W/q$.

An additional
non trivial check of the validity of (B16) is available when
$v_{\rm int}=0$ in (B2).  In this case as shown in section IIIB, the neutral
sector may be solved exactly by mapping onto chiral fermions.
We have checked that the ensemble averaged Green's function
$G_{\sigma\sigma}$
calculated from this exact solution agrees with the form
in (B16).

{}From (B15) we see that for $W=0$ there are two propagating
modes and, using the results of appendix A, the
conductance is non universal and given by (2.19).  In contrast,
for any finite $W$, in the limit $q << W^{-1}$, there is only
a single propagating mode $\omega = (v_\rho - v_{\rm int}^2/v_\sigma)q$.
This reflects the fact that at finite temperatures,
when there is interchannel tunneling
present, only the total charge is conserved, so that there is a
single propagating mode.  Moreover, it may be explicitly verified
from (B15) that the conductance is given by the quantized value
$G = \nu e^2/h$.

We now consider the analogous calculation in the vicinity of the
SU(n) random fixed point.  In this case, as argued in
section IIIC, we wish to compute the self energy for $G_1$, defined
in (3.21), due to the random perturbation $v_{12}$, which has
mean square average $W_c$ and couples
to $\partial_x\phi_\rho \exp i\sqrt{2}\chi_1$.  The self energy
to leading order in $W_c$ is computed by evaluating the
diagrams in figure 3.  We find
\begin{equation}
\Sigma(q,\omega_n) =
W_c \int_0^\beta d\tau (e^{i\omega_n\tau} - 1) P(\tau)
\end{equation}
where,
\begin{equation}
P(\tau) = < T[\partial_x\phi_\rho(\tau)e^{i\chi_1(\tau)}
\partial_x\phi_\rho(0) e^{-i\chi_1(0)}]>
\end{equation}
\begin{equation}
 \propto \left[ {\pi/\beta} \over \sin (\pi\tau/\beta) \right]^4.
\end{equation}
Upon analytically continuing to real frequency, we thus find
\begin{equation}
\Sigma(q,\omega) = {i\omega\over{2\pi \ell_\sigma} }
\end{equation}
with
$\ell_\sigma \propto W_c T^2$.
It may again be checked that terms in the self energy higher order
in $W_c$ are singular as $q\rightarrow 0$.  Using arguments analogous
to those presented above, we conclude that to correctly describe the
long wavelength physics, we must replace (B20) by
\begin{equation}
\Sigma (q,\omega) = {i\omega\over{2\pi\ell_\sigma}}
{1\over{1 - i(q\ell)^{-1}}}  .
\end{equation}

\begin{table}
\begin{tabular}{ccccc}
$\nu$       & $1/3$ & $2/5$ & $3/7$ & $4/9$  \\ \hline
$\alpha$    &  $4$  & $4$   & $4$   & $4$    \\ \hline\hline
$\nu$       & $2/3$ & $3/5$ & $4/7$ & $5/9$  \\ \hline
$\alpha$    & $2$   & $8/3$ & $3$   & $31/5$ \\
\end{tabular}

\caption{Tunneling exponent $\alpha$ for the
temperature dependent conductance (1.2) through
a point contact separating two quantum Hall fluids at filling
factor $\nu$.}
\end{table}

\begin{figure}
\caption{Schematic portrait of the edge of a quantum Hall state
with two channels.  The presence of random impurities, denoted
by the small circles, allows for momentum non-conserving scattering
between the different channels.  When the channels move in the
same direction (e.g. $\nu=2$), as shown in (a),
inter-channel scattering does not effect the net transmission of
the edge.
However, when the channels move
in opposite directions, as in $\nu=2/3$, depicted in (b), the back
scattering of charge plays a crucial role.}
\end{figure}

\begin{figure}
\caption{Schematic diagram of a two terminal conductance measurement
for a quantum Hall state with two channels which move in opposite
directions (i.e. $\nu = 2/(2p+1)$ with $p<0$).
The shaded regions denote the reservoirs.}
\end{figure}

\begin{figure}
\caption{Diagrams for the self energy of
the Greens function in (3.21).  The solid
circles represent the interaction (3.19).  The solid lines
represents the bare propagator for $\chi_1$, $G_1^0$, and
a sum over all possible combinations of these lines is implied.
The wavy line represents the bare propagator for $\phi_\rho$.}
\end{figure}

\begin{figure}
\caption{Schematic portrait of a point contact, in which the top and
bottom edges are brought together by an electrostatically controlled gate,
allowing for the tunneling of charge between the two
edges. }
\end{figure}

\begin{figure}
\caption{Diagrams for the self energy
of the Greens function in (B4).
The solid circles represent the tunneling operator (B3).
The solid lines represent the bare propagator for $\phi_\sigma$,
$G^0_{\sigma\sigma}$.  A sum over all possible combinations of
these lines is implied.}
\end{figure}


\begin{references}

\bibitem{lutt1}
J.M. Luttinger J. Math. Phys., {\bf 15}, 609 (1963).

\bibitem{lutt2}
A. Luther and L.J. Peschel, Phys. Rev. B {\bf 9} 2911 (1974);
Phys. Rev.
Lett. {\bf 32}, 992 (1974);
A. Luther and V.J. Emery, Phys. Rev. Lett. {\bf 33}, 589 (1974).

\bibitem{lutt3}
J. Solyom, Advances in Physics {\bf 28}, 201 (1970);
V.J. Emery in {\it Highly Conducting One-Dimensional Solids},
edited by J.T. Devreese (Plenum Press, New York 1979).

\bibitem{Haldane1}
F.D.M. Haldane, J. Phys. C {\bf 14}, 2585 (1981);
F.D.M. Haldane, Phys. Rev. Lett. {\bf 47}, 1840 (1981).

\bibitem{Wind}
U. Meirav, M.A. Kastner, M. Heiblum and S.J. Wind, Phys. Rev. B
{\bf 40}, 5871 (1989).

\bibitem{wire}
G. Timp, in {\it Mesoscopic Phenomena in Solids},
edited by B. L. Altshuler, P.A. Lee and R.A. Webb (Elsevier, Amsterdam, 1990).

\bibitem{Webbp}
R. Webb, private communication (1993).

\bibitem{review}
See, for example, C.W. J. Beenakker and H. van Houten,
in {\it Solid State Physics}, edited by H. Ehrenreich
and D. Turnbul (Academic, New York, 1991), Vol. 44.

\bibitem{Wen1}
X.G. Wen, Phys. Rev. B {\bf 43}, 11025 (1991); Phys. Rev. Lett.
{\bf 64}, 2206 (1990).  X.G. Wen, Phys. Rev. B {\bf 44} 5708 (1991).


\bibitem{MacDonald}
A. H. MacDonald, Phys. Rev. Lett. {\bf 64}, 222
(1990); M. D. Johnson and A. H. MacDonald, Phys. Rev. Lett. {\bf 67},
2060 (1991).

\bibitem{Moon}
K. Moon et. al., Phys. Rev. Lett. {\bf 71}, 4381 (1993).

\bibitem{Prange-Girvin}
See for example, {\it The Quantum Hall effect}, edited by R. Prange and S.M.
Girvin,
(Springer-Verlag, New York, 1990).

\bibitem{Halperin1}
B.I. Halperin, Phys. Rev. B {\bf 25}, 2185 (1982).

\bibitem{Buttiker}
M. Buettiker, Phys. Rev. Lett. {\bf 57}, 1761 (1986).

\bibitem{Wen2}
''Impurity effects on chiral 1D electron systems", X. Wen,
preprint (1994).

\bibitem{Alphenaar}
B.W. Alphenaar, P.L. McEuen, R.G. Wheeler and
R.N. Sacks, Phys. Rev. Lett. {\bf 64}, 677 (1990).

\bibitem{Kouwenhoven}
L.P. Kouwenhoven, {\it et. al.} Phys. Rev. Lett. {\bf 64}, 685 (1990).


\bibitem{Webb}
F. Milliken, C. Umbach and R. Webb, "Evidence for
Luttinger liquid in the Fractional quantum Hall regime", IBM preprint (1994).

\bibitem{HaldaneHalperin}
F.D.M. Haldane, Phys. Rev. Lett. {\bf 51}, 605 (1983);
B.I. Halperin, Phys. Rev. Lett. {\bf 52}, 1583 (1984).

\bibitem{us23}
C. L. Kane, M.P.A. Fisher and J. Polchinski,
Phys. Rev. Lett. {\bf72}, 4129 (1994).

\bibitem{Jain}
J.K. Jain, Phys. Rev. Lett. {\bf 63}, 199 (1989).

\bibitem{Read}
N. Read, Phys. Rev. Lett.  {\bf 65}, 1502 (1990).

\bibitem{WenZee}
See X.G. Wen and A. Zee, Phys. Rev. B {\bf 46}, 2290 (1992),
and references therein.

\bibitem{GL}
S.C. Zhang, T.H. Hansson and S. Kivelson, Phys. Rev. Lett.
{\bf62}, 82 (1989);  N.Read, Phys. Rev. Lett.
{\bf62}, 86 (1989);  S.M. Girvin and A.H. MacDonald,
Phys. Rev. lett. {\bf58}, 1252  (1987).

\bibitem{coulomb}
A long range Coulomb interaction leads to a wavevector
dependent velocity in (2.7), $v \propto \ln(1/q)$.
It can be shown that this will restore the quantization of
the conductance in an infinite sample.  However, in a finite
sample of size $L$, there will be non negligible corrections to the
quantized conductance which are of order $\ln^{-2} L/a$, where
$a$ is the magnetic length.


\bibitem{us1}
C.L. Kane and M.P.A. Fisher,
Phys. Rev. B {\bf 46}, 15233 (1992).

\bibitem{Zeesun}
J. Frohlich and A.Zee, Nuclear Physics B{\bf 364}, 517 (1991).

\bibitem{Ashoori}
R.C. Ashoori, H. Stormer, L. Pfeiffer, K. Baldwin and K. West,
Phys. Rev. B {\bf 45}, 3894 (1992).

\bibitem{Giamarchi}
A similar perturbative RG is described in detail in,
T. Giamarchi and H. Schulz, Phys. Rev.B {\bf37}, 325 (1988).

\end{references}
\end{document}